\documentclass[twocolumn]{aa}
\usepackage{graphicx}
\usepackage{txfonts}
\begin{document}

\title{Two spectral states of the transient X-ray burster SAX~J1747.0-2853}

\titlerunning{Two spectral states of the transient X-ray burster SAX~J1747.0-2853}
\author{L. Natalucci\inst{1}, A. Bazzano\inst{1}, M. Cocchi\inst{1}, P. Ubertini\inst{1},
        R. Cornelisse\inst{2,3,4}, J. Heise\inst{2,3}, J.J.M. in~'t Zand\inst{2,3}
	}

\authorrunning{Natalucci et al.}

\offprints{L. Natalucci}

\institute{CNR-Istituto di Astrofisica Spaziale e Fisica Cosmica,
           Area Ricerca Roma 2/Tor Vergata, Via del Fosso del Cavaliere 100,
           00133 Roma, Italy\\
\email{lorenzo@rm.iasf.cnr.it}
              \and
            SRON National Institute for Space Research, Sorbonnelaan 2, NL-3584 
	    CA Utrecht, the Netherlands\\
              \and
	    Astronomical Institute, Utrecht University, P.O.Box 80000, NL-3508 TA
	    Utrecht, the Netherlands \\
	      \and
	    Dept. of Physics and Astronomy, University of Southampton, Hampshire
	    SO17~1BJ, U.K.
       }

\date{Received; Accepted 04/12/2003 }


\abstract{
The neutron star binary SAX~J1747.0-2853, located in the Galactic Center
region at about 0.5 deg from Sgr~A* and at a distance of $\sim9$~kpc,
has been observed in outburst four times (1998, 1999, 2000 and 2001) 
by {\em BeppoSAX} and {\em RossiXTE}.
At the time of its discovery in 1998 the source was observed in a low/hard state,
showing a hard tail with a high energy cutoff of $\sim70$~keV. About two 
years later the source reappeared about one order of magnitude brighter in
the X-rays (0.5-10 keV) and with a significantly steeper spectrum. 
As was the case for the low state, the data could be fitted by an input model 
based on two continuum primary components: 
a) a soft thermal excess, which is 
$\sim4$ times more luminous than the one found in hard state; b) a
non-thermal component which is compatible with either a power-law or a 
comptonization spectrum. The soft component  
is equally well described by pure blackbody or multi-color disk emission,
with significantly higher temperature than observed in low state 
($\sim1.3$ vs. the $\sim0.5$~keV assuming pure blackbody).
For this model, the flux of the non-thermal component below $\sim10$~keV is 
a significant fraction of the total X-ray flux, i.e. greater than $\sim50$\% 
in the 2-10 keV band. 
}

\maketitle

\section{Introduction}

X-ray bursters are an important and rich subclass of low-mass
X-ray binaries (LMXB), consisting of weakly magnetized neutron stars
accreting from a low mass companion and experiencing thermonuclear flashes.
These flashes, normally lasting a few seconds to minutes, are known in the 
literature as type~I bursts (see reviews in \cite{Lew93}; \cite{Str03}) and 
produce mainly X-rays by unstable burning of (a fraction of)
the material accreted on the surface of the neutron 
star (NS). 
Sometimes very energetic X-ray bursts are observed, characterized by an
expansion of the neutron star atmosphere, due to an energy release   
above the Eddington limit. These can be easily identified
as the expansion of the photosphere and the subsequent cooling 
strongly modify the spectrum and the temporal profile of the burst. 
The determination of the maximum intensity can be used to estimate 
the distance to the source on the assumption of burst emission isotropy, 
of given object mass (usually the standard 1.4~${M}_{\odot}$ for a NS) 
and helium abundance (e.g., \cite{Kuu03}).

The transient source SAX~J1747.0-2853, discovered in March 1998 by the 
Wide Field Cameras on board {\em BeppoSAX}  
(\cite{Zan98}) has been seen to emit many X-ray bursts, including a 
radius expansion event which allowed an estimate of the source distance 
as $\sim9$ kpc (\cite{Nat00}; hereafter, Paper~I). As the source has a 
small angular distance 
to the Galactic Center (GC), at 0.5~deg from Sgr~A*, it is very likely that
it is located deep in the GC region, probably at a distance of no more 
than a few hundred parsecs from Sgr~A*. During the discovery outburst 
lasting $\approx3$~weeks (see Paper~I), a wide band spectrum was 
obtained by observation with the {\em BeppoSAX} Narrow Field Instruments 
(NFI), and was found to be dominated 
by a hard tail extending at least up to $\sim200$~keV. The source showed
a slow rise, low intensity outburst, and reached a luminosity of  
$\sim3\times10^{36}$~erg~s$^{-1}$ (2-10 keV) at its maximum.
The continuum emission was well fitted by a Comptonization spectrum  
(\cite{Tit94}) plus
a soft component which could be well described by a blackbody spectrum
with temperature $kT$=($0.55\pm0.10$)~keV. 
Two years later, in the spring of 2000 a new, brighter outburst
was detected with the source intensity reaching about 140 mCrab
(\cite{Mar00}; \cite{Mur00}). This is the most luminous outburst 
detected so far from this source. A {\em RXTE}/PCA light curve was published 
by in~'t Zand et al. (2001) showing a fast rise, exponential decay light 
curve with multiple peaks.
The {\em BeppoSAX}/NFI were pointed again in the direction of SAX~J1747.0-2853 on 
2000 March 13, when the source intensity was $\sim100$~mCrab in the 2-10
keV band ($\sim1.8\times10^{37}$~erg~s$^{-1}$), and 
the source had just started to decline from its maximum. The results of this 
observation will be described in the present paper.

\begin{figure}
\centering
\includegraphics[width=8.5cm,angle=-90]{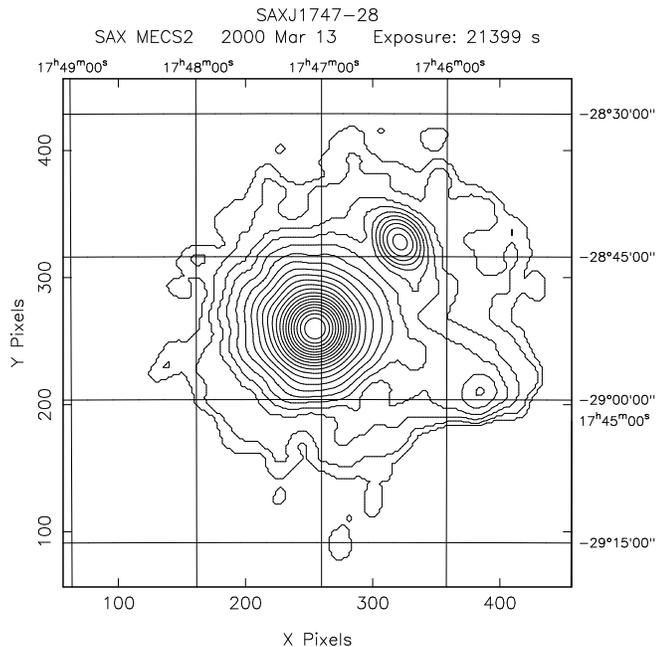}
\caption{
MECS detector image of the field including SAX~J1747.0-2853, taken
on 2000, March 13. Contour levels are logarithmically scaled. Two other
sources are visible in the field-of-view: 1E1743.1-2843 on the upper
right, and Sgr~A* (lower right). 
      }
\label{Fig1}
\end{figure}

After the March 2000 outburst, SAX~J1747.0-2853 showed activity for more   
than one year, with detection of more bursts by the
{\em BeppoSAX}/WFC and occurrence of 
minor outbursts (\cite{Wer03}). A significant, brighter flare
took place on Sept 2001 with a peak intensity of $\sim60$~mCrab.
as shown by the {\em RXTE}/ASM light curve
published by Wijnands, Miller \& Wang (2002). 
During this activity period,  
the source was observed by {\em Chandra} in its Galactic Center 
Survey on July 8th, 2001. It was detected with an unabsorbed flux
of $\sim1$~mCrab in the X-ray band (Wijnands, Miller \& Wang, 2002), 
i.e. a 0.5-10 keV luminosity  
of $3\times10^{35}$~erg~s$^{-1}$. The {\em Chandra} observation allowed to 
determine with high accuracy the source position:
$\alpha=17^{h}$~$47^{m}$~$02.604^{s}$ and 
 $\delta=-28^{\circ}$~$52$'~$58.9$"
(equinox 2000, 1$\sigma$ error radius $\sim0.7\arcsec$). A new 
determination of the source distance is available by the detection 
of a radius expansion burst by the {\em GRANAT}/ART-P telescope 
on October 20, 1991 (\cite{Gre02}). The authors reported 
a distance of $(7.9\pm0.4)$~kpc assuming a 1.4~${M}_{\odot}$ NS. This
is fully consistent with the value of $(8.9\pm1.3)$~kpc reported in 
Paper~I, obtained from spectral analysis of the 1998 radius expansion burst. 

The detection of the early burst by ART-P is intriguing. An upper limit to
the source luminosity at the epoch of the burst is set by Grebenev et al. 
(2002) at $1.1\times10^{36}$~erg~s$^{-1}$ in the 4-26 keV band. 
This suggests that the source was active more than ten years ago, 
and that the occurrence of regular, frequent outbursts is a  
characteristic of SAX~J$1747.0-2853$.     
This is also supported by the possibility of earlier detections with 
error boxes consistent with the position of the source, namely, the transient
GX~0.2-0.2 observed by {\em Ariel~5} (Proctor, Skinner \& Willmore 1978).

\section{Observations and data analysis}

The {\em BeppoSAX}/NFI were pointed towards SAX~J1747.0-2853 on 2000 March 13, for
a single observation with a total elapsed time of $5.7\times10^{4}$~s. An image 
obtained with one of the units of the Medium Energy Concentrator Spectrometer 
(MECS, \cite{Boe97}) is shown in Fig~\ref{Fig1}. The MECS active detectors for this
observations were the two units MECS2 and MECS3, for a total exposure time
of $2.1\times10^{4}$~s.

\begin{table*}[]
\caption[]{
Spectral model fits to the
emission spectrum of SAX~J1747.0-2853 in the energy range 0.2-10.5~keV. 
Model names follow the same notation as in XSPEC v.11. All quoted errors 
refer to deviations at 90\% confidence. ${N}_{H}$ is the Wisconsin
absorption parameter. For fits with comptonization model, the plasma 
temperature and optical depth are largely unconstrained and therefore are
not reported (see text for more details). \\
Parameters description and units: 
${N}_{H}$, Wisconsin absorption parameter in units of
 $10^{22}$~cm$^{-2}$;
 $\Gamma$, photon index of power law component;
 ${kT}_{0}$, temperature of the comptonized soft seed photons, in keV;
 ${kT}_{bb}$, temperature of the blackbody component, in keV;
 ${R^2}_{bb}$, normalization expressed as $r^2/d^2$,
where r is the blackbody radius in km and d is the distance to the source
in units of 10 kpc;
${R^2}_{in}$~$\cos{\theta}$, where ${\theta}$ is the
 disk inclination angle and ${R}_{in}$ is the disk inner radius in km,
for a source at 10 kpc;
 ${E}_{K\alpha}$, centroid energy of gaussian line profile, in keV.
}
\vspace{-1mm}
\begin{flushleft}
\begin{tabular}{l|c|l|c}
\hline\hline 
  
\vspace{-3mm}\\
  Model & ${N}_{H}$ & Parameter values and errors& $\chi^2_\nu$[dof]  \\
  & ($10^{22}$~cm$^{-2}$) &  &  \\
\hline

\vspace{-3mm}\\
BB+Diskbb+Gaussian  & $9.03^{+0.27}_{-0.16}$ &
 ${kT}_{in}$=$1.05^{+0.22}_{-0.10}$, ${R^2}_{in}$~$\cos{\theta}$=$144^{+90}_{-78}$,
 ${kT}_{bb}$=$1.58^{+0.18}_{-0.04}$, ${R^2}_{bb}$=$33\pm11$, &
 1.08 [110] \\
 & & ${E}_{K\alpha}$=$6.96\pm0.09$ & \\
BB+Powerlaw+Gaussian  & $10.89^{+0.40}_{-0.32}$ &
  ${kT}_{bb}$=$1.32\pm0.05$,
  ${R^2}_{bb}$=$50.4^{+6.2}_{-4.8}$, $\Gamma=3.14^{+0.19}_{-0.14}$,
  ${E}_{K\alpha}$=$6.97\pm0.09$ &
 1.05 [110] \\
Diskbb+Powerlaw+Gaussian  & $10.87^{+0.34}_{-1.03}$ &
 ${kT}_{in}$=$1.92^{+0.02}_{-0.10}$,
 ${R^2}_{in}$~$\cos{\theta}$=$13.7\pm1.1$, $\Gamma$=$3.70^{+0.45}_{-0.63}$,
 ${E}_{K\alpha}$=$6.96\pm0.09$  &
 1.03 [110] \\
BB+Comptt+Gaussian  & $8.77^{+0.62}_{-0.50}$ &
 ${kT}_{bb}$=$1.33\pm0.08$,  ${R^2}_{bb}$=$54.3^{+5.8}_{-7.5}$, 
 ${kT}_{0}$=$0.50^{+0.08}_{-0.04}$, ${E}_{K\alpha}$=$6.93^{+0.04}_{-0.14}$  &
 1.03 [108] \\
Diskbb+Comptt+Gaussian & $9.08^{+1.20}_{-0.43}$ &
 ${kT}_{in}$=$1.89^{+0.07}_{-0.12}$, ${R^2}_{in}$~$\cos{\theta}$=$15.3^{+2.6}_{-1.0}$, 
 ${kT}_{0}$=$0.44^{+0.16}_{-0.28}$, ${E}_{K\alpha}$=$6.93^{+0.07}_{-0.02}$ &
 1.05 [108] \\

\hline
\end{tabular}
\end{flushleft}
\label{tab:spectral}
\end{table*}

\begin{figure}
\centering
\includegraphics[width=5.6cm,angle=-90]{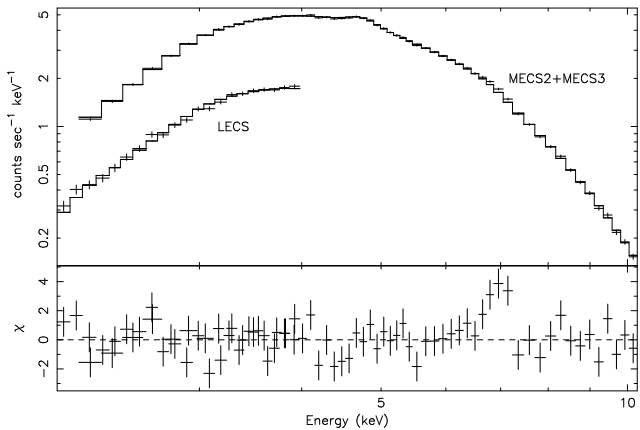}
\caption{The count rate spectrum of the combined concentrator instruments,
for the March 2000 observation, in the energy range 2-10.5 keV. Data points
are shown for the LECS (thin crosses), and two combined MECS units 
(thick crosses).
The model shown is the best fit to a blackbody plus power-law continuum.
The residuals show a weak iron emission line near 7 keV, fully compatible with 
the MECS energy resolution.
      }
\label{Fig2}
\end{figure}

Being very close to GC, the observed field is complex. In the MECS images,
two other sources are visible in the field-of view: 1E1743.1-2843 
(Cremonesi et al., 1999; Pourquet et al., 2003) and the X-ray source associated 
with Sgr~A*. At the epoch of the observation, HPGSPC was
no more operational, so the energy bands available were:
the 0.2-10.5 keV range, covered by the Low Energy Concentrator Spectrometer (LECS, 
Parmar et al. 1997) and MECS; and 15-200 keV, covered by the Phoswich Detector
System (PDS, \cite{Fro97a}). The LECS exposure time for this observation
was $7.8\times10^{3}$~s. For LECS and MECS we used the following energy bands  
for analysis: 0.2-4.0 keV for LECS and 1.8-10.5 keV for MECS. At high
energies, problems of source contamination led us to exclude the PDS data
from the spectral fitting: in particular, the high energy source
1E~1740.7-2942 is at $1.06\deg$ from the pointing position, resulting in a 
($15\pm5$)~\% illumination of the detector area, as estimated from the 
PDS collimator response (\cite{Fro97b}). This contamination 
was much less important for the 1998 observation, due to both the strenght
of the hard X-ray emission from SAX~J1747.0-2853 and its hard spectral 
shape (see discussion in Paper~I). The same considerations, however, are not true 
for our March 2000 observation, because the total high energy flux
detected by the PDS was lower by a factor $\sim2$ and also, the measured
spectrum is much steeper (see discussion below). 

We then decided to use only low energy data for spectral analysis. To 
subtract the background from the concentrator's data, we used the 
procedure described in Paper~I, with the same background
extraction regions. We carefully checked the MECS light curve to verify
that no X-ray bursts were present. 
In Table~1 are listed the results of the spectral fitting. All the best fit
models include a weak iron emission line, which is detected above the
continuum with an equivalent width of ($33\pm2)$~eV (see also Fig.~2). The 
agreement between data and models is high ($\chi^{2}_{r}\leq1.08$ for all the fits,
without including any systematic errors). The line is narrow and its centroid is 
$\approx6.95$~keV. Without this narrow feature, the quality of the fit degrades 
to a $\chi^{2}_{r}\approx1.3$ for $\approx105$~{\em dof}, so we consider 
its detection significant. An iron line of similar strenght and energy 
was also observed in March 1998, in the {\em low/hard state}, together with 
an absorption edge at $\approx7.4$~keV. No indication of an absorption edge 
is present in the 2000 data, when adding this component in all model fits. 
There are basically two classes of models that can adequately describe
the observed spectrum. The first one is a completely thermal spectrum, for 
which we have found an acceptable fit by adding two components: a pure 
blackbody spectrum, characterized by $kT\sim1.6$~keV and an emission radius 
compatible with the NS size, and a multi-color disk (MCD) blackbody 
(\cite{Mit84}) with an inner disk temperature ${kT}_{in}$~$\approx1$~keV. 
The second class of models include a power-law or comptonized component, plus a 
blackbody (yielding $kT\sim1.3$~keV) or MCD blackbody. 

In both model classes, the maximum of the emitted power is found in the
band 2-10~keV (see Fig.~3). However, if the comptonized component is present 
(as for the case of the 1998 low state) there would be significant 
hard X-ray emission. In the case of thermal Comptonization (i.e., {\em comptt}
in XSPEC v.11) the shape of the hard X-ray spectrum depends on the 
Comptonization parameter $y$ (\cite{Tit94}): for a spherical cloud geometry,
$y=3\tau_{0}^{2}$~${kT}_{e}$~/~${m}_{e}c^{2}\pi^{2}$. 
The spectral fits below 10.5~keV cannot constrain
the plasma temperature ${kT}_{e}$ and the optical depth $\tau_{0}$,
but only the seed photon temperature ${kT}_{0}\sim0.4$~keV (corresponding to the  
original energy spectrum of the scattered photons) which is not relevant to
determine the high energy tail. The best fit Comptonization
spectra reported in Table~1 converge at ${kT}_{e}\sim15$~keV. These yield,
in the energy band 2-10~keV, an overall (unabsorbed) flux of 
$1.95\times10^{-9}$~erg~cm$^{-2}$~s$^{-1}$
of the comptonized component against a corresponding value of 
$1.57\times10^{-9}$~erg~cm$^{-2}$~s$^{-1}$ of the blackbody flux. 

In order to investigate the presence of hard X-ray emission above $\sim30$~keV
we have analyzed the PDS spectrum measured during the observation. Indeed,
all the models listed in Table~1 are significantly below the
spectral points of the PDS. Fitting the latter with a power-law, yields a 
photon index $\Gamma=2.81\pm0.14$ ($\chi^{2}_{r}=1.2$ for 9 d.o.f.) in the 
energy range 30-200~keV, for a total measured intensity of $\sim40$~mCrab. 
The spectral index is too high to be totally ascribed to 
illumination by 1E~1740.7-2942 (see e.g. \cite{Mai99}). The
model flux at $\sim200$~keV is about 10 times lower than the spectrum 
obtained by Sidoli et al. (1999) for 1E~1740.7-2942, which is compatible
with the expected factor in exposed area for the two spectra. However, 
this ratio reduces to about a factor 3 at $\sim50$~keV. This suggests 
that additional high energy flux is present at $E\leq100$~keV.  
Later observations of the same field have detected significant variability of 
the PDS flux (\cite{Wer03}) proving that most of the high energy flux 
comes from point sources. Besides 1E~1740.7-2942, the only known high 
energy source located close to our transient (at $\sim40\arcmin$) is the 
persistent X-ray burster A1742-294. So far, the latter has been detected  
above $\sim30$~keV only once, with an intensity of 
$\approx30$~mCrab in the 35-100~keV band (\cite{Chu95}). For these reasons, 
it is very likely that a significant fraction of the high energy flux in
the range 30-100~keV is originated from SAX~J1747.0-2853. In this case, the 
first class of models based on thermal only (NS and disk) origin would be
ruled out for this observation.

\section{Conclusions}

\begin{figure}
\centering
\includegraphics[width=9.0cm]{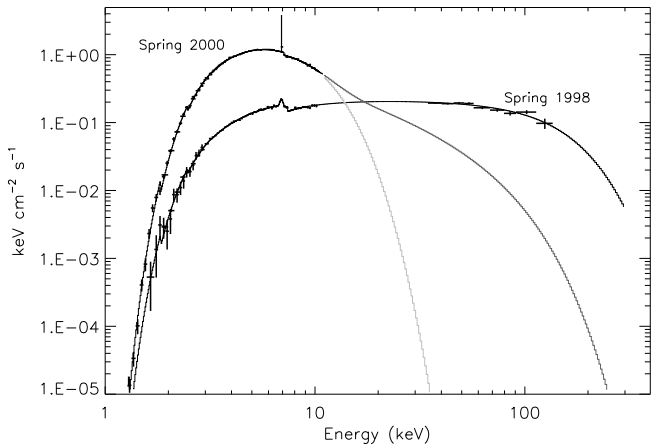}
\caption{
    The unfolded spectra measured by the {\em BeppoSAX}/NFI showing two different
    spectral states of SAX~J1747.0-2853. The best model spectra derived from the 
    two observations of 1998 and 2000 are also shown in
    the energy range 1-200~keV. For the March 2000 observation, where valid spectral 
    measurements are available only below 10.5~keV, the high energy extrapolation
    is shown for two models, the first including a possible hard tail produced by 
    thermal comptonization  
    (dark grey), the second entirely thermal with a multi-color disk and a 
    blackbody component (light grey). 
      }
\label{Fig3}
\end{figure}

The two observations performed by {\em BeppoSAX} in 1998 and 2000 show
the existence of two different spectral states for the transient burster 
SAX~J1747.0-2853 (see Fig.~3).
In March 2000, the prominent hard tail observed in the 1998 {\em low/hard}
state outburst is absent or significantly reduced, whereas the X-ray
emission is characterized by a soft component with temperature $kT\approx1.3$~keV.
This is a factor $\approx4$ higher than the one detected in
the {\em low/hard} state, with a significant change in temperature
($kT\approx0.5$~keV in the low state spectrum). The emission
region of the thermal component, determined from the observed {\em colour} 
temperature, has a size corresponding to a blackbody radius ${R}_{bb}\sim7$~km.
i.e. coincident or very close to the NS surface. This component, when 
modelled by MCD, has an inner disk temperature of 
$\approx1.9$~keV and a normalization corresponding to an inner radius
${R}_{in}$~$\approx4.5$~/~$\sqrt{\cos{\theta}}$~km, where $\theta$ is the
angle of inclination. We remark that, for both models, the measured radii 
are most probably underestimated by a factor $\approx2$, due to the 
expected hardening of the original thermal spectrum induced by electron 
scattering (\cite{Ebi94}, \cite{Shi95}). 

The X-ray spectrum seems to be characterized by a significant non-thermal component 
which could be responsible for a substantial fraction of the total X-ray flux. 
Using as reference the 2-10 keV range, the detected intensity of this
component is greater than $\sim2\times10^{-9}$~erg~cm$^{-2}$~s$^{-1}$, i.e.
$\geq50$\% of the total. The production of hard X-rays during the rise
and close to the maximum of an outburst is a feature commonly observed in 
transient Black Hole Candidates (BHC). This is the so-called 
{\em very high} state (see e.g., \cite{Che96}), to distinguish it from the 
{\em high} state which occurs later, and is characterized by the luminosity of 
the non-thermal component being comparable to the disk luminosity. The spectrum 
of SAX~J1747.0-2853 seems highly reminiscent of this behaviour. Nevertheless, 
we could not firmly establish the characteristics of the non-thermal component 
by direct spectral measurements at $E\geq10$~keV, due to a high source 
confusion being present in this region. 

As for the 1998 observation, the X-ray spectrum measured in the spring of 2000
presents evidence for a weak (equivalent width $\approx33$~eV), narrow emission 
line near $\approx7$~keV. The energy of both lines suggests a ionization stage 
compatible with Fe~XXVI. An iron absorption edge has been detected only 
in the 1998 spectrum at ($7.45\pm0.15$)~keV. Its energy indicates a 
significantly lower ionization. It is then possible that the iron line originates 
from a fully ionized portion of the accretion disk (probably the inner region) 
or in the NS boundary layer, whilst the edge observed in 1998 is produced in a 
different, mildly ionized absorbing region of the disk. The Fe edge 
is not detected in the {\em high/soft} state.  

\begin{acknowledgements}
The {\em BeppoSAX} satellite is a joint Italian and 
Dutch programme. We thank ASI for continuous support of the 
mission and operations.  
\end{acknowledgements}
 

\end{document}